# Relative stereociliary motion in a hair bundle opposes amplification at distortion frequencies

Andrei S. Kozlov,[1] Thomas Risler,[2,3,4] Armin J. Hinterwirth,[1] and A. J. Hudspeth[1]

[1]Howard Hughes Medical Institute and Laboratory of Sensory Neuroscience, The Rockefeller University, New York, New York; [2]Institut Curie, Centre de Recherche, F-75005, Paris, France; [3]UPMC Université Paris 06, UMR 168, F-75005, Paris, France; [4]CNRS, UMR 168, F-75005, Paris, France

## NON-TECHNICAL SUMMARY

The hair cell, or sensory receptor of the inner ear, achieves high sensitivity by amplifying its mechanical inputs. The mechanism of amplification depends on the concerted opening and closing of mechanically sensitive ion channels in the hair bundle, a cluster of actin-containing rods that protrude from the cell's top surface. When a hair cell is stimulated simultaneously at two frequencies, channel gating also produces distortion products or responses at other frequencies. Using a sensitive interferometer to measure the motions of stereocilia, we have found that hydrodynamic forces act within the hair bundle to suppress these spurious signals. The hair bundle has evidently evolved an effective means of amplifying input signals while reducing the effect of distortions.

## ABSTRACT

Direct gating of mechanoelectrical-transduction channels by mechanical force is a basic feature of hair cells that assures fast transduction and underpins the mechanical amplification of acoustic inputs. But the associated nonlinearity—the gating compliance—inevitably distorts signals. Because reducing distortion would make the ear a better detector, we sought mechanisms with that effect. Mimicking *in vivo* stimulation, we used stiff probes to displace individual hair bundles at physiological amplitudes and measured the coherence and phase of the relative stereociliary motions with a dual-beam differential interferometer. Although stereocilia moved coherently and in phase at the stimulus frequencies, large phase lags at the frequencies of the internally generated distortion products indicated dissipative relative motions. Tip links engaged these relative modes and decreased the coherence in both stimulated and free hair bundles. These results show that a hair bundle breaks into a highly dissipative serial arrangement of stereocilia at distortion frequencies, precluding their amplification.

**ABBREVIATIONS:** DIC, differential interference contrast; HEPES, 4-(2-hydroxyethyl)-1-piperazineethanesulfonic acid; RMS, root mean square

## INTRODUCTION

The sensitivity and frequency selectivity of hearing are controlled by the physics of mechanoelectrical transduction in the ear's receptor, the hair cell. Each hair cell bears on its apical surface a bundle of cylindrical stereocilia separated by small gaps filled with viscous endolymph. Stereocilia are connected to their closest neighbors by several protein linkages. Among these, the tip links connect the tip of each stereocilium to the side of its tallest neighbor and translate hair-bundle deflection into tension that controls the open probability of mechanosensitive transduction channels (Hudspeth, 1989). Coherent motion of the stereocilia in a hair bundle minimizes viscous energy dissipation and preserves at the cellular level the mechanical nonlinearity that originates from individual channel molecules in the form of gating compliance (Howard and Hudspeth, 1988; Kozlov et al., 2011). Coupled to a source of energy through myosin motors, gating compliance contributes to an active process that lowers the threshold of sound detection by counteracting the dissipative influence of the surrounding liquid (reviewed in Hudspeth, 1997; Fettiplace et al., 2001; LeMasurier and Gillespie, 2005; Fettiplace and Hackney, 2006; Hudspeth, 2008). Together, these features make the hair bundle a highly sensitive mechanical detector and amplifier.

A negative consequence of direct channel gating is that the ear distorts and sometimes suppresses external sounds. In particular, the ear can produce new vibrations at frequencies that correspond to integer combinations of those present in the stimulus. Such internally generated distortion products interact with external signals at other locations along the cochlea, creating additional distortion products that interact further with the input. The result is a cascade of oscillations, all internally generated by the ear and absent from the original sound (Robles et al., 1997; Jülicher et al., 2001). The cause of these distortions is the nonlinear response that originates in the direct gating of the transduction channels by mechanical force.

Unidirectional amplification, in which the active movements of hair bundles do not feed back onto the basilar membrane, can in principle decrease distortion in a mammalian ear poised at a Hopf bifurcation (Reichenbach and Hudspeth, 2011). Using a protocol that mimics *in vivo* stimulation, we have investigated the production of distortion products to ascertain whether a hair bundle can amplify stimuli while precluding amplification of the distortions.

**METHODS**

*Experimental preparation*

Using a protocol approved by the Institutional Animal Care and Use Committee of The Rockefeller University, we performed experiments on sacculi from adult bullfrogs (*Rana catesbeiana*) of both sexes. After having been immersed in water at 0 °C, each animal was doubly pitched and decapitated, a procedure considered acceptable by the Panel on Euthanasia of the American Veterinary Medical Association (www.avma.org/issues/animal_welfare/euthanasia.pdf.). The sacculi were then dissected as described previously (Kozlov et al., 2007) and maintained in oxygenated saline solution comprising 120 mM NaCl, 2 mM KCl, 1 mM $CaCl_2$, 10 mM D-glucose, and 5 mM 4-(2-hydroxyethyl)-1-piperazineethanesulfonic acid (HEPES) at pH 7.3.

Following a 30-60 min digestion at room temperature with $1 \ mg \cdot ml^{-1}$ collagenase (type XI, Sigma Chemical Co.), the otolithic membrane was removed and the sensory epithelium of the sacculus was peeled from the underlying connective tissue. The epithelium was then folded and secured in an experimental chamber by a golden electron-microscopic grid. All experiments were performed at a room temperature of 20 °C to 25 °C in the saline solutions used during dissection.

*Stimulation*

We stimulated each hair bundle with a glass pipette about 1 μm in internal tip diameter and attached only to the kinociliary bulb. A larger pipette attached to the tall edge of a hair bundle entirely suppressed the distortion products there (Kozlov et al., 2011), but that was undesirable in this study for two reasons. First, *in vivo* only the kinociliary bulb is attached to the otolithic or tectorial membrane and displaced, so our experimental arrangement resembled the physiological configuration. Second, to characterize the phase angles at the distortion frequencies we required that these distortions be present at both edges of a hair bundle.

***Dual-beam differential interferometer***

A dual-beam interferometer allowed the detection of picometer-scale movements by transparent microscopic structures. Based on an earlier single-beam design (Denk et al., 1989; Denk and Webb, 1990), it featured two independently controlled beams, each of which acted as an interferometer (Fig. 1). A conventional upright microscope with differential-interference-contrast (DIC) capability (BX51WI, Olympus) served as the basis of the system. The two interferometers were kept independent by assigning a different wavelength to each: a 633 nm, 1.8 mW red He-Ne gas laser (117A, Spectra-Physics) provided one interferometric beam and a 532 nm, 10 mW green diode-pumped solid-state laser (85 GCA 010, Melles Griot) the other.

For each laser, an optical isolator (IO-3D-λ-PHE, OFR Thorlabs Inc.) prevented intensity oscillations resulting from back-reflections into the laser cavity. The light intensity could be modified as necessary with a compensated attenuator (925B, Newport). The beam profile was spatially filtered by coupling it to a single-mode optical fiber. At the fiber's exit port, a single-mode fiber coupler (F-91-C1, Newport) provided a parallel beam of a diameter matching the aperture of the microscope's condenser lens. Passing the beam through a polarizer and a quarter-wave plate rendered it circularly polarized before it entered the microscope.

The beams shared the second lens of a beam-steering telescope. The first components of this steering assembly were positioned such that second lens imaged them in the condenser's aperture plane. This arrangement rendered a translation of the steering lenses into a translation of the lasers' focal points in the specimen plane. Each steering lens could be moved either manually with a micrometer-controlled linear stage for coarse positioning of the focal spot or with a controllable piezoelectrical actuator (P-263, PiezoJena) for fine positioning and calibration.

The two parallel beams were directed into the microscope by a dichroic mirror situated below the condenser. At this position infrared light from a conventional halogen lamp was combined with the laser beams to illuminate the specimen. To facilitate positioning of the beams on a hair cell, the infrared image was acquired by a charge-coupled-device camera and displayed on a video monitor.

We used two identical 40x water-immersion objectives of numerical aperture 0.8 (LUMPlanFI/IR, Olympus) in a symmetrical configuration to provide both a condenser that focused the beams to a diffraction-limited spot in the specimen plane and an observation lens. The condenser stage of the microscope was modified to accommodate the objective lens as well as a DIC prism (WI-DICT2, Olympus) located in its aperture plane. The lower DIC prism split each circularly polarized beam into two sub-beams that were polarized orthogonally to each other and spatially separated by about 100 nm at the specimen. The sub-beams were then recombined in a second, identical DIC prism situated above the objective lens. If both sub-beams encountered optical paths of the same length on their passage through the specimen, their recombination yielded a circularly polarized beam. If one of the sub-beams instead experienced an optical path of a different length as a result of variations in refractive index or specimen thickness, an elliptically polarized beam ensued after recombination.

To measure the ellipticities of the red and green laser beams, each was isolated with a dichroic mirror and passed through a narrow bandpass filter transmissive for the appropriate wavelength. This arrangement ensured that there was no measurable cross-talk between the two channels. Each polarized beam was then split into two orthogonal components by a polarizing beam splitter. Photodiodes attached to transimpedance amplifiers with a roll-off frequency of about 20 kHz reported the intensity of each E-vector component. The analog signals were filtered with eight-pole Butterworth filters with a low-pass corner frequency of 20 kHz, then digitized with 16-bit resolution at sampling intervals of 10 μs. A computer calculated the final deflection values from the detector output for each channel by taking the ratio of the difference between the two E-vector components to their sum. This ratio was directly related to the phase difference between the two sub-beams, which in turn depended on the position of the object that the two sub-beams traversed.

During the initial calibration, pixel values of the video system were converted into distances by use of a stage micrometer with 10-μm increments. A further calibration step related the voltage applied to the piezoelectric actuator that moved either beam's steering lens with the ensuing movement of the respective focal spot. To measure the distance of movement, the focal spot was imaged with the video system and its centroid determined (Image Processing Toolbox, MATLAB). This centroid was tracked during the calibration procedure and used as the position coordinate of the focal spot.

### Data acquisition

A single measurement consisted of a set of twenty records, each lasting either 0.1 s or 1 s. Each record corresponded to a double time series, acquired simultaneously and independently with the two laser beams directed at desired positions on a hair cell. Occasionally, when small debris in the optical path or laser pointing instabilities corrupted the signal, a record was rejected on the basis of an unstable root-mean-square amplitude, an abnormal cross-correlation amplitude, or a pronounced drift in the raw time series.

Calibration was achieved during an experiment by translating the focal spot of each laser beam across the specimen by a known distance while recording the resulting detector signals. Because the calibration factor between focal-spot movement and detector output depended on exactly where the beam was positioned, the detection apparatus was calibrated independently for each beam before and after each record.

### Spectral analysis

In order to characterize the degree of common and relative stereociliary motion within the hair bundle, we measured simultaneously the time-dependent positions of stereocilia at two different locations in a hair bundle. From these measurements, we derived the spectral coherency function, a normalized, complex-valued quantity. The modulus, called the coherence, is a real number between zero and one that characterizes the degree to which two objects move together at each frequency. The phase characterizes the delay between the two time series at each frequency. For each of these quantities, spectral estimates were derived by the Thomson multitaper method with six orthogonal tapers and confidence intervals were obtained using jackknife estimation methods (Thomson, 1982; Thomson and Chave, 1991; Percival and Walden, 1993). The justification for this method and further details can be found in the Supplementary Information.

A central assumption underlying spectral-estimation methods is that the time series under consideration are stationary. Nonstationarities in the data were controlled by three means. First, we acquired relatively short time series between which the measurement apparatus was recalibrated to compensate for slow drift. This approach provided a set of independent records that could be used to estimate the desired quantities by an approach similar to overlapped-segment averaging. Next, our criteria for outlier rejection ensured that strongly nonstationary records were eliminated. Finally, the multitaper spectral-estimation technique prevented low-frequency artifacts from contaminating the estimated spectral densities.

### *Statistical estimates over different records and measurements*

For each measurement in this study, we selected a variable number $B$ of records and further analyzed them with $K = 6$ orthogonal Slepian data tapers resulting from the choice of a bandwidth parameter $W$ such that $2NW\tau = 8$ for data series of length $N$ and sampled at a rate $\tau$. Such a choice ensured $K \leq 2NW\tau$ and that the spectral power of the tapers was sufficiently concentrated in the frequency band $\left[ f_0 - W, f_0 + W \right]$ (Supplementary Information). For statistical purposes, the resulting $P = K \times B$ spectral estimates were considered as independent and equivalent estimates of the same stochastic process. The jackknife variance estimate was then performed on this entire set of realizations.

For each experimental condition, several measurements apiece were performed on multiple cells. The spectral estimate associated with the set of results corresponding to each experimental condition is presented as the mean ± standard deviation for each frequency.

## RESULTS

Using the whole-epithelium preparation from the bullfrog's sacculus (Kozlov et al., 2007), we examined the coherency of stereociliary motion in mechanically excited hair bundles using the dual-beam differential interferometer (Fig. 1). To stimulate each hair bundle, we used a glass pipette attached by suction to the kinociliary bulb. The stiff glass probe was displaced with a piezoelectrical device simultaneously at 90 Hz and 115 Hz, a procedure that evokes distortion products at well-defined frequencies (Jaramillo et al., 1993; Kozlov et al., 2011). We used displacements comparable to those that occur *in vivo* in response to strong physiological stimuli, typically 50-100 nm in root-mean-square (RMS) magnitude.

We examined the magnitude and phase of the relative motions between stereocilia adjacent to the stimulus probe and those on the opposite side of the hair bundles both at the frequencies of stimulation and at those of the internally generated distortion products. Six stimulated hair bundles displayed a coherence of essentially unity and a phase of approximately zero at the frequencies of stimulation: the coherence and phase at 90 Hz were 0.9992 ± 0.0005 and -0.023 ± 0.034 rad, and those at 115 Hz were 0.9992 ± 0.0005 and -0.027 ± 0.047 rad (Fig. 2). In contrast, at the distortion-product frequencies the coherence dropped sharply and the motion developed a pronounced phase lag. For example, at the distortion frequency $f_2 + f_1 = 205$ Hz the coherence decreased to 0.51 ± 0.19 and the phase lag was as large as -1.65 ± 2.08 rad.

In principle, differences in signal amplitude can artificially affect the coherence by changing the signal-to-noise ratio when noise unrelated to hair-bundle motions is present. A comparison of the coherence and spectral power at the distortion frequencies with those at immediately adjacent frequencies corresponding to the background noise shows that the reduced coherence of the distortion components is associated with an increased spectral power (Fig. 2). Because a poor signal-to-noise ratio would produce the opposite correlation, this result indicates that the drop of coherence and the large phase lags at the distortion frequencies are not noise artifacts but are characteristics of the hair-bundle kinematics. In addition, a poor signal-to-noise should yield phases that are ill-defined but of zero mean. That we observe instead large mean phases at the distortion frequencies buttresses the previous affirmation. Such results in the coherence and phase spectra indicate the presence of highly dissipative modes of motion at the distortion components (Kozlov et al., 2011). Note that the phase lags at the various distortion frequencies (Table 1) were correlated with the amount of force that the hair bundle exerted against the stimulating probe at those frequencies (Jaramillo et al., 1993).

Forces in the tip links, including those originating from channel gating, can change the separation between adjacent stereocilia. Although these relative movements are tiny, they can have a dramatic effect on energy dissipation by exciting modes of motion with high viscous drag (Kozlov et al., 2011). The decrease in coherence observed at the distortion frequencies suggested that it might also be possible to detect a drop caused by the tip links in unstimulated hair bundles undergoing free fluctuations. Because tip links connect stereocilia only along the axis of mechanosensitivity, we compared the coherence for a hair bundle measured along this axis with that obtained in the orthogonal direction. By performing experiments on hair cells confirmed to have functional mechanotransduction by the presence of spontaneous oscillations, we selected hair bundles with intact tip links.

In control experiments on 17 hair bundles, the coherence between 100 Hz and 5 kHz for the same-stereocilium measurements was $0.97 \pm 0.04$. When measured at the opposite sides of 29 hair bundles under identical conditions, the coherence fell to $0.92 \pm 0.05$ (Fig. 3). For movements in the direction orthogonal to the axis of mechanosensitivity, the average coherence in this frequency range for the opposite sides of seven hair bundles was $0.97 \pm 0.02$ (Fig. 3) *versus* 0.98 ± 0.01 for same-stereocilium measurements in the identical cells. In every instance the mean phase value was close to zero at all frequencies.

The comparison shows that the coherence is higher for motions in the orthogonal plane than for those in the plane of mechanosensitivity ($P = 0.0048$ by a two-tailed *t*-test). Therefore, even though tip links elastically connect adjacent stereocilia, they globally disorder stereociliary motion in a hair bundle on a sub-nanometer scale. This seemingly paradoxical observation accords with quantitative predictions of a detailed finite-element model of the bullfrog's saccular hair bundle (Kozlov et al., 2011).

The magnitude of thermal movement in the direction orthogonal to the axis of mechanosensitivity for these seven cells, 9.6 ± 1.5 nm RMS, exceeded that in the direction of mechanosensitivity in the 29 cells confirmed to have intact mechanotransduction, 7.3 ± 3.2 nm ($P = 0.08$). To test whether these differences in signal amplitude could have artificially affected the coherence, we investigated whether there was any significant correlation between the coherence and magnitude values. Correlation coefficients were insignificant for both the cells observed along the axis of mechanosensitivity (correlation coefficient -0.002, $P = 0.99$) and those observed in the orthogonal direction (correlation coefficient -0.1, $P = 0.83$). This further validates the comparison of the coherence between the two orientation axes of a hair bundle.

## DISCUSSION

A lack of sufficiently accurate experiments has fostered confusion about the degree of relative stereociliary motion and the roles of interstereociliary linkages and the surrounding liquid in coupling stereocilia. Theoretical analyses imply that the thermal fluctuations of stereocilia with differing lengths display distinct spectral characteristics (Svrcek-Seiler et al., 1998). In accordance with this idea, modeling studies suggest that the tension in tip links is not evenly distributed across a hair bundle, but instead decreases along a stereociliary file (Duncan and Grant, 1997; Cotton and Grant, 2004; Nam et al., 2007a, 2007b). These results notwithstanding, morphometric analysis indicates that the stereocilia are tightly packed against one another at rest, without separation along the axis of mechanosensitivity (Jacobs and Hudspeth, 1990). Microscopic observations and video measurements disclose that stereocilia do not separate significantly when hair bundles are subjected to large, low-frequency stimuli (Hudspeth and Corey, 1977; Karavitaki and Corey, 2010) and even to rapid displacement steps (Crawford et al., 1989; but see Duncan et al., 1999). Furthermore, the stiffness of a hair bundle is proportional to the number of its stereocilia, which suggests an equal distribution of force among them (Howard and Ashmore, 1986). Although interstereociliary connections are commonly thought to transmit mechanical forces across a hair bundle and to allow it to move coherently (Goodyear and Richardson, 1999, Nam et al., 2006), fluid-dynamical calculations emphasize the importance of the viscous interactions between stereocilia and the liquid (Zetes and Steele, 1997). The disagreements among these studies stem from the difficulty of resolving very small amplitudes of stereociliary movement and from the uncertainties intrinsic to modeling mechanically complex, many-body systems such as the hair bundle.

Measurements obtained with a dual-beam interferometer indicate that the relative motions in hair bundles are tiny compared to the common motion over a broad range of frequencies. This result obtains both when the hair bundles fluctuate freely (Kozlov et al., 2007) and at the stimulus frequencies when they are excited mechanically (Kozlov et al., 2011). Although morphological differences between the hair bundles of mammals and those of other vertebrates could lead to quantitatively different results, modeling of the stereociliary motion in mammalian bundles suggests a similar behavior (Kozlov et al., 2011). Considering the advantages conferred to the ear by the coherent motion of stereocilia, it would not be surprising if such a mode of operation is a general feature of all hair bundles. Indeed, we found that auditory hair bundles in the tokay gecko also show highly concerted motion of their stereocilia (Supplementary Information).

We have demonstrated that hair bundles display a different behavior at the frequencies of the internally generated distortion products, for which squeezing modes of motion between adjacent stereocilia become significant. This can be seen from the presence at these frequencies of strongly non-zero mean phases indicative of the anti-phase components in the responses. The principal cause of this kinematic behavior is geometrical: because tip links are oriented obliquely to the long axes of the stereocilia, they pull adjacent stereocilia toward one another when tension increases and allow them to separate when it decreases. Even in unstimulated cells this decreases the coherence in the direction of mechanosensitivity as compared with that in the orthogonal direction. Because the relative motions occur at sub-nanometer amplitudes, they are not readily detectable by diffraction-limited imaging techniques (Karavitaki and Corey, 2010).

The amplitudes of the relative modes of stereociliary motion decrease with frequency owing to the viscous coupling between stereocilia (Kozlov et al., 2011). At sufficiently high frequencies, typically in the kilohertz range, the tip links should not be able to cause relative motions and to lower the slope of the displacement-open probability relation at the distortion frequencies. Consistent with this idea, a cascade of distortion products occurs at the cochlea's high-frequency base (Robles et al., 1997). At the cochlear apex, however, where the nonlinearities are believed to be milder (Robles and Ruggero, 2001), the internal damping described here could diminish the amount of distortion.

In a stimulated hair bundle the common mode of motion completely dominates any relative mode at the stimulus frequencies. In contrast, at the distortion frequencies, internally generated squeezing modes are of major importance. Although the active process can counteract the rather modest viscous drag for the common mode at the stimulus frequencies, at the distortion frequencies the hair bundle breaks into a serial arrangement of stereocilia with pronounced squeezing modes that have a thousand-fold higher drag compared to that of the common mode (Kozlov et al., 2011). Amplification can operate at the frequencies of stimulation at which the bundle moves coherently, but should be impaired at the distortion frequencies by independent motions of the stereocilia that decrease the slope of the relation between displacement and open probability and thus limit the sensitivity of mechanoelectrical transduction. It follows that clustering stereocilia in a hair bundle provides a double advantage to the ear. First, it suppresses drag at the stimulus frequencies and thus facilitates amplification by the active process (Kozlov et al., 2011). And second, it increases the drag and decreases the coherence at the frequencies of the distortion products, thereby opposing their amplification. The ear therefore takes advantage of the direct gating of mechanotransduction channels by mechanical forces, including the nonlinear amplification that this feature affords, and at the same time diminishes the distortions created by these nonlinearities.

**ACKNOWLEDGMENTS**: We thank B. Fabella for technical support and programming and O. Ahmad, D. Andor, J. Baumgart, and M. O. Magnasco for discussions. This investigation was funded by grant DC000241 from the National Institutes of Health. A. S. K. was supported by Howard Hughes Medical Institute, of which A. J. Hudspeth is an Investigator.

**AUTHOR CONTRIBUTIONS**: A. S. K. and A. J. Hudspeth conceived the project; A. S. K. and A. J. Hinterwirth constructed the interferometer; A. S. K. performed the experiments; T. R. implemented the data analysis; A. S. K. and T. R. analyzed the data and wrote the manuscript; A. J. Hinterwirth and A. J. Hudspeth edited the manuscript. The experiments were conducted at The Rockefeller University.

Table 1: Coherence and phase values at the stimulation and principal distortion-product frequencies.

| Frequency (Hz) | Coherence ± SD | Phase (radians) ± SD |
|---|---|---|
| 90 ($f_1$) | 0.99923 ± 0.00051 | −0.0230 ± 0.0340 |
| 115 ($f_2$) | 0.99922 ± 0.00050 | − 0.0272 ± 0.0478 |
| 180 ($2 \cdot f_1$) | 0.32 ± 0.15 | −0.97 ± 2.46 |
| 205 ($f_1 + f_2$) | 0.51 ± 0.19 | −1.65 ± 2.08 |
| 230 ($2 \cdot f_2$) | 0.42 ± 0.20 | −0.54 ± 1.08 |

**FIGURES**

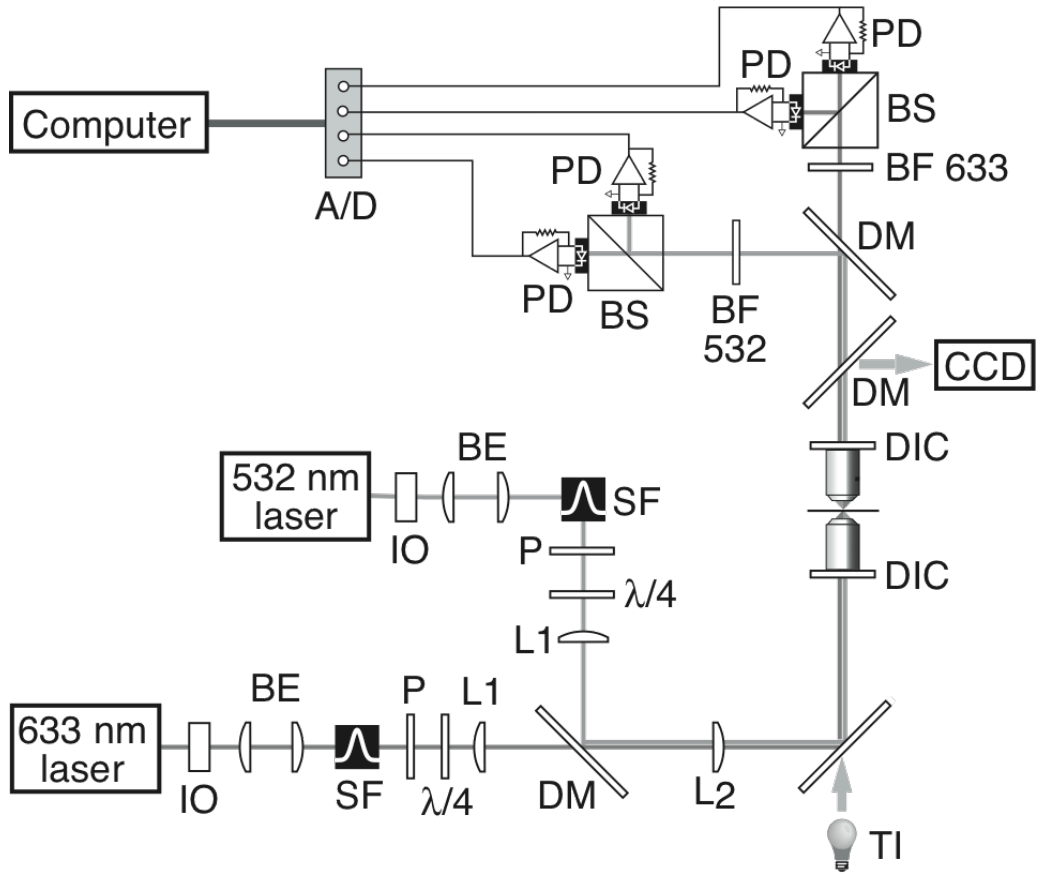

**Figure 1: Schematic diagram of the double-beam differential interferometer.** Two lasers provide independent beams of coherent light. The abbreviations signify: IO, optical isolator; BE, beam expander; SF, spatial filter with a single-mode optical fiber; P, polarizer; $\lambda/4$, quarterwave plate; L1, steering lens; DM, dichroic mirror; L2, fixed lens; TI, tungsten illuminator; DIC, differential-interference contrast system including objective lens and Wollaston prism; CCD, charge-coupled-device camera; BF, bandpass filter; BS, beam splitter; PD, photodiode with current-to-voltage converter and amplifier; A/D, analog-to-digital converter.

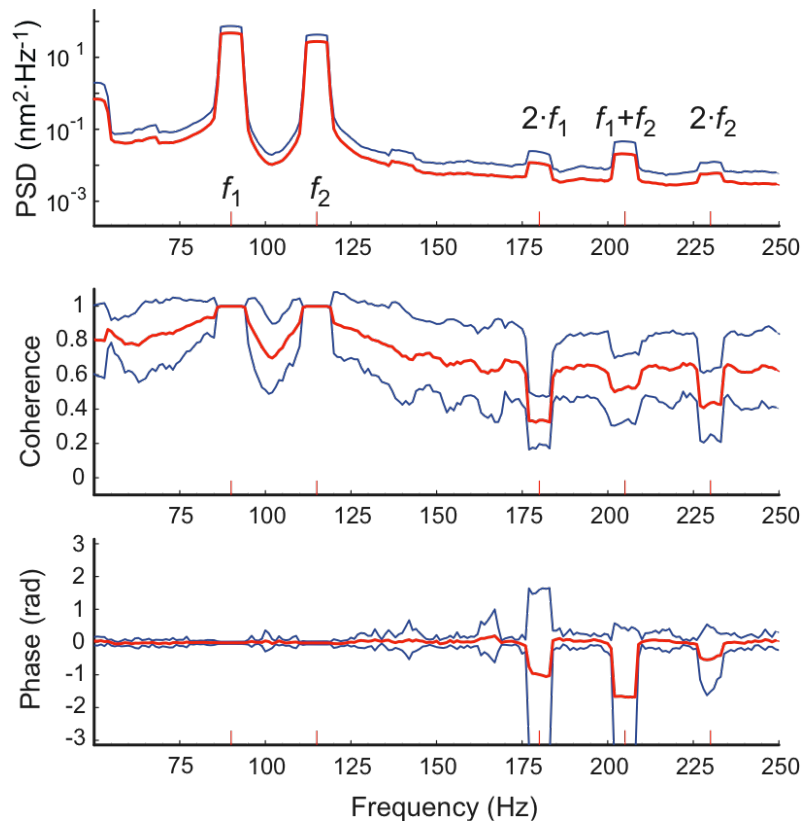

**Figure 2: Spectral power, coherence and phase at the distortion frequencies.** The average values of the spectral-power density (PSD), coherence and phase for six stimulated cells are shown in red and the standard deviations in blue. Because of the use of a logarithmic scale for the spectral-power density, the values corresponding to the mean minus a standard deviation are not shown in this case. Stereociliary movements at the frequencies of stimulation, $f_1 = 90$ Hz and $f_2 = 115$ Hz, display a coherence across the hair bundle close to one and a phase close to zero; at those frequencies the bundle moves as a unit. In contrast, large phase lags and coherence drops are apparent at the distortion frequencies $2 \cdot f_1 = 180$ Hz, $f_2 + f_1 = 205$ Hz, and $2 \cdot f_2 = 230$ Hz. The relatively large associated power-spectral densities confirm that the observed phase lags are not due to a poor signal-to-noise ratio at these frequencies. Although smaller drops in coherence are also apparent at other distortion frequencies, they are typically less prominent than those at the quadratic distortion frequencies, in agreement with earlier results (Jaramillo et al., 1993; Kozlov et al., 2011).

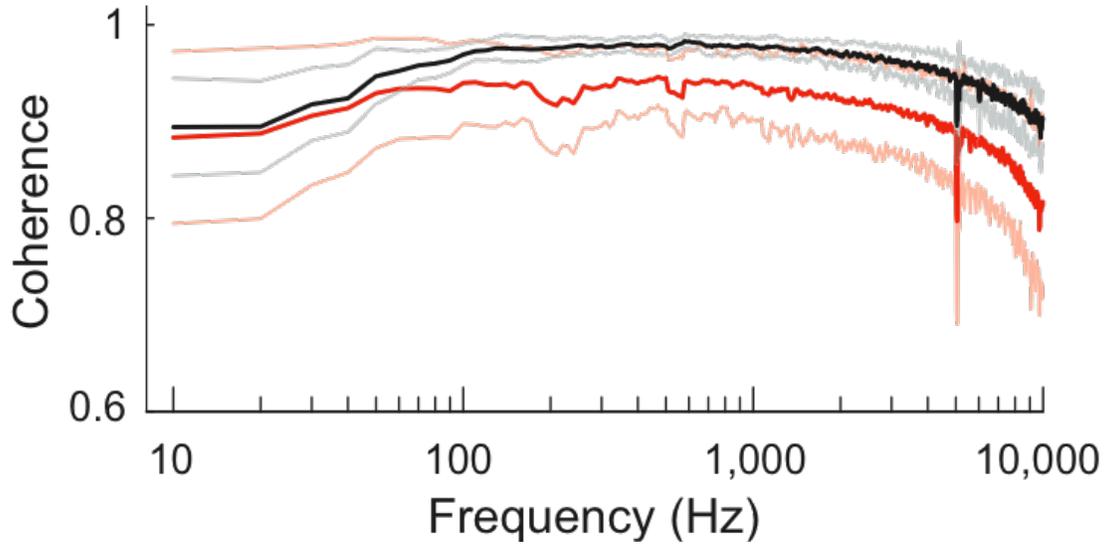

**Figure 3**: **Coherence along the two perpendicular axes of hair-bundle movement.** The average coherence values for movement in the orthogonal plane, which are shown in black, exceed those for movement in the plane of mechanosensitivity, which are portrayed in red. The associated standard deviations are displayed in grey and pink, respectively. In both configurations the mean phase was around zero, but the standard deviation between 100 Hz and 5 kHz was smaller for movement in the orthogonal plane than for movement in the plane of mechanosensitivity—0.04 rad *versus* 0.16 rad—a result that accords with the greater coherence in the former configuration.

**SUPPLEMENTARY INFORMATION**

# Relative stereociliary motion in a hair bundle opposes amplification at distortion frequencies

Andrei S. Kozlov,[1] Thomas Risler,[2,3,4] Armin J. Hinterwirth,[1] and A. J. Hudspeth[1]

[1]Howard Hughes Medical Institute and Laboratory of Sensory Neuroscience, The Rockefeller University, New York, New York; [2]Institut Curie, Centre de Recherche, F-75005, Paris, France; [3]UPMC Université Paris 06, UMR 168, F-75005, Paris, France; [4]CNRS, UMR 168, F-75005, Paris, France

The experiments presented in the main text were performed on hair cells from the bullfrog's sacculus, an organ responsive to frequencies around 100 Hz (Lewis, 1988). We wished to test the generality of our previous conclusion that hair cells display high coherence (Kozlov et al., 2007) in auditory hair bundles tuned to higher frequencies. For this purpose we recorded from hair bundles of the basilar papilla in the tokay gecko (*Gekko gecko*), a lizard with a well developed sense of hearing that extends to about 7 kHz (Manley, 1972).

Cochleae dissected from geckos (Chiappe et al., 2007) were maintained in oxygenated saline solution comprising 170 mM NaCl, 2 mM KCl, 1 mM $CaCl_2$, 10 mM D-glucose, and 5 mM HEPES at pH 7.3. After a 30-60 min digestion of basilar papillae at room temperature in 1 mg/ml collagenase (type XI, Sigma Chemical Co.), hair cells were mechanically isolated with an eyelash and allowed to settle onto the bottom of an experimental chamber coated with either concanavalin A (Sigma) or Cell-Tak (BD Biosciences) to promote cellular adhesion.

A typical pair of time series obtained from the opposite edges of a gecko's hair bundle demonstrated fluctuations with a root-mean square (RMS) magnitude of $3.5 \pm 0.5$ nm (mean ± standard deviation; Fig. S1A). The associated spectra showed a high coherence across the hair bundle with a phase difference near zero at all frequencies (Fig. S1B). A complete measurement, which comprised twenty such records, yielded similar coherence and phase spectra (Fig. S1C). These results did not differ significantly from those obtained with the two laser beams superimposed at the same position on the hair bundle (Fig. S1D). The average coherence and phase spectra obtained from five hair bundles confirm the consistency of the results (Fig. S2).

Measurements of coherency from isolated hair cells may be corrupted by whole-cell drift. We therefore performed on each isolated cell a control measurement in which one laser beam was positioned on the hair bundle and the other on the apical portion of the cell body. If the two signals displayed cross-correlations above the background noise level, the cell was rejected from further analysis. For the five cells included in the analysis, the average coherence between measurements from cell bodies and the associated hair bundles was $0.2 \pm 0.1$ and the average phase was $0.0 \pm 1.7$ rad between 100 Hz and 5 kHz. This result is consistent with the expectation for two independent signals. In contrast, when the two laser beams were positioned on the opposite sides of these hair bundles, the coherence was $0.88 \pm 0.07$ and the phase was $0.0 \pm 0.1$ rad (Fig. S2). These values indicate that auditory hair bundles from the tokay gecko move with a high coherence.

Experimental studies of hair-bundle kinematics with high resolution in isolated cells suffer from at least three methodological drawbacks: cell isolation may cause excessive mechanical and metabolic damage; the proximity of a hair bundle to the recording chamber's glass surface may affect the hydrodynamic forces experienced by the stereocilia; and isolated cells may drift with the local hydrodynamic flow. Although drifting could be strongly reduced by covering the recording chamber with adhesive substances such as concanavalin A and Cell-Tak, most cells nevertheless displayed whole-cell motions of several tens of nanometers. Because these movements were large compared with those typical of relative stereociliary motions, they dominated the measurements. When their global motion was negligible, we could use some isolated cells for recordings of hair-bundle motion. However, stimulating isolated cells with a patch pipette attached to their hair bundles displaced the cell bodies too much to allow reliable measurements of the relative stereociliary motion. Because the basilar papilla of the tokay gecko is thin, with only a dozen hair cells abreast, we could not fold it to provide optical access to laterally protruding hair bundles without damaging the cells. For these reasons, we report in the main text the results from our experiments with the bullfrog's sacculus, which has the advantage of being mechanically stable and metabolically robust: we occasionally observed hair bundles oscillating spontaneously for as long as 20 hours *in vitro*.

### *Choice of the multitaper spectral-analysis technique*

The structure of the data was investigated through linear frequency-domain representations using the multitaper spectral-estimation method (Thomson, 1982; Percival and Walden, 1993). This choice is motivated by the observation that standard spectral-estimation methods based on single data windows suffer from two fundamental problems, namely bias and lack of consistency of the so-obtained estimates. The first problem refers to the fact that the estimate of the spectral quantity at a given frequency mixes information from different frequency components of the original signal. The second problem refers to the fact that the variance of the estimate does not decline with an increasing sample size, for the outcome contains as many quantities as there are data values. Data tapering by a single window, which is often used in an attempt to solve the first problem, suffers from variance-efficiency reduction, unequal weighting of the data, and arbitrariness in taper selection (Brillinger, 1981; Thomson, 1982). In our case, the bias reduction offered by the use of a single window was also inadequate because our records contained a relatively important part of low-frequency signal that was unrelated to the desired observations and that could have contaminated the frequency components of interest. To solve the second problem, a convolution product in the frequency domain could be used to smooth the desired estimates, but this operation relies on the assumption that the output spectral quantities are smooth. In our case the spectral power was concentrated at specific frequencies and this condition was unsatisfied.

*Details of the multitaper spectral analysis*

Any stationary stochastic process $X(t)$ sampled at a rate $\tau$ can be characterized by its Cramer spectral representation

$$X(t) = \int_{-1/(2\tau)}^{1/(2\tau)} dX(f) e^{2\pi i f t} \tag{1}$$

for any time $t$ at which it is sampled. Here $dX(f)$ is an orthogonal-increment process: for zero-mean processes $E\{dX(f)\} = 0$; for distinct frequencies $f_1$ and $f_2$, $dX(f_1)$ and the complex conjugate of $dX(f_2)$ are statistically uncorrelated. The second moment of this function defines the power spectrum $S(f)$ of the process,

$$S(f)df = E\left\{\left|dX(f)\right|^2\right\}. \tag{2}$$

In actual experiments, however, we can observe only a specific realization $x(t)$ of $X(t)$ and can do so only over the finite time window $T = N\tau$. The observed time series $\tilde{x}(t)$ has a Fourier transform $\tilde{x}(f)$ that is related to the Fourier transform $x(f)$ of the infinite time series $x(t)$ by

$$\tilde{x}(f) = \sum_{j=1}^{N} x(t_j) e^{-2\pi i f t_j} = \int_{-1/(2\tau)}^{1/(2\tau)} K(f - f', N) x(f') df', \tag{3}$$

in which $t_j = j\tau$ and

$$K(f) = e^{-\pi i f(N+1)} \left[\frac{\sin(N\pi f)}{\sin(\pi f)}\right]. \tag{4}$$

For a stationary stochastic process the spectrum can be estimated as $\left|\tilde{x}(f)\right|^2$, the squared Fourier transform of the data series. However, $\tilde{x}(f)$ is not equal to $x(f)$ but is related to it by a convolution product that mixes information originating in different frequency channels (Equation 3); this corresponds to the first problem of bias mentioned above. Moreover, $\left|\tilde{x}(f)\right|^2$ squares the observations without averaging them, estimating $N$ quantities from $N$ data values. The resultant over-fitting problem corresponds to the second problem mentioned above, namely lack of consistency.

The multitaper spectral-estimation method provides an elegant solution to both problems (Thomson, 1982; Percival and Walden, 1993). In this approach, the data are multiplied not by a single window but rather by a set of $K$ optimally chosen data tapers $w_k(t)$. The power spectrum is then estimated as

$$S_{\mathrm{MT}}(f) = \frac{1}{K} \sum_{k=1}^{K} \left| \tilde{x}_k(f) \right|^2,$$  (5)

in which

$$\tilde{x}_k(f) = \sum_{j=1}^{N} w_k(t_j) x(t_j) e^{-2\pi i f t_j} .$$  (6)

The optimal choice of the taper functions requires that they be mutually orthogonal, providing $K$ independent spectral estimates, and that they possess maximal spectral concentration, yielding the greatest relative power over the frequency bandwidth $2W$ (Mitra and Pesaran, 1999). The spectral concentration value is quantified by

$$\lambda_k(N,W) = \frac{\int_{-W}^{W} \left| U_k(f) \right|^2 \mathrm{d}f}{\int_{-1/(2\tau)}^{1/(2\tau)} \left| U_k(f) \right|^2 \mathrm{d}f} ,$$  (7)

in which $U_k(f)$ is the Fourier transform of the sequence $w_k(t)$ and $\tau$ is the sampling rate. It can be shown that $\lambda_k(N,W)$ is the $k^{th}$ eigenvalue of the eigenvector relation

$$\sum_{j'=1}^{N} \frac{\sin\left[ 2\pi W (t_j - t_{j'}) \right]}{\pi(t_j - t_{j'})} w(t_{j'}) = \lambda w(t_j);$$  (8)

$w_k(t)$ corresponds to the associated eigenvector under appropriate normalization.

These optimal taper functions have the remarkable property that $2NW\tau$ of their eigenvalues are approximately equal to one, whereas the remainder decay sharply to zero (Slepian and Pollak, 1961; Slepian, 1978). As a result, the spectrum is convolved with a window that is as close as possible to a rectangular shape, yielding a power-spectral estimate that results at each frequency $f_0$ from an equally weighted average of contributions across the frequency band $\left[ f_0 - W, f_0 + W \right]$. At the same time, leakage to and from frequencies outside this frequency band is as small as possible.

We chose as the final spectral estimate an equally weighted average of the different tapered spectra. Although more sophisticated techniques of averaging can be implemented with both frequency- and data-dependent weighting (Thomson, 1982; Percival and Walden, 1993; Mitra and Pesaran, 1999), these approaches significantly affect the results only when tapers with poor spectral concentration are used. Because we could average over several independent records for each measurement, we did not need to include a large number of tapers to obtain consistent estimates. We could therefore restrict ourselves to tapers with high concentration properties, rendering adaptive-weighting techniques superfluous.

### *Jackknife error estimation*

The multitaper method also permits calculation of the variance of an estimated spectrum by means of jackknifing (Thomson and Chave, 1991). Although this method is applicable to relatively complicated data, it is largely free of distributional assumptions and hence highly reliable. Moreover, the jackknife variance always exceeds the true variance, so the estimates are conservative (Efron and Stein, 1981).

The jackknife approach was implemented as follows. Let $\{x_i\}$, $i = 1,...,P$ be a sample of $P$ independent observations drawn from some distribution characterized by a parameter $\theta$ to be estimated, and let $\hat{\theta}$ be an estimate of $\theta$. In addition to the usual estimate $\hat{\theta}_{\text{all}}$ based on all $P$ observations, we formed $P$ estimates $\{\hat{\theta}_{\backslash i}\}$, each based on the $(P-1)$ observations remaining after deletion of the $i^{\text{th}}$ one. In the nonparametric estimation of the variance of an arbitrary statistics,

$$\text{var}\left(\hat{\theta}\right) = \frac{P-1}{P}\sum_{i=1}^{P}\left(\hat{\theta}_{\backslash i} - \theta_{\cap}\right)^2, \tag{9}$$

in which

$$\theta_{\cap} = \frac{1}{P}\sum_{i=1}^{P}\hat{\theta}_{\backslash i}. \tag{10}$$

These estimates require the use of transformations prior to jackknifing when the statistics is bounded or its distribution is strongly non-Gaussian, for example strongly asymmetrical about the mean value. We used the logarithmic transformation for jackknifing power spectra and the inverse hyperbolic transformation for jackknifing coherences (Thomson and Chave, 1991).

In power-spectral estimation, a logarithmic transformation stabilizes the distribution into a more symmetric one. Taking $\hat{\theta} = \ln \hat{S}$ as our estimator, we formed the delete-one values $\ln \hat{S}_{\backslash i}$ as

$$\ln \hat{S}_{\backslash i} = \ln \left( \frac{1}{P-1} \sum_{\substack{k=1 \\ k \neq i}}^{P} \hat{S}_k \right), \tag{11}$$

with their average

$$\ln S_{\cap} = \frac{1}{P} \sum_{i=1}^{P} \ln \hat{S}_{\backslash i} \tag{12}$$

defining a power-spectral estimate $S_{\cap}$. It follows that the jackknife estimate of the variance of the logarithmic power spectrum was

$$\hat{\sigma}^2 = \text{var} \left\{ \ln \hat{S} \right\} = \frac{P-1}{P} \sum_{i=1}^{P} \left( \ln \hat{S}_{\backslash i} - \ln S_{\cap} \right)^2. \tag{13}$$

Because of the logarithmic transformation, $\left( \ln \hat{S}_{\backslash i} - \ln S_{\cap} \right) / \hat{\sigma}$ was distributed nearly as $t_{P-1}$, a $t$-distribution with $P-1$ degrees of freedom. The approximate $(1-\alpha)$ confidence interval for the power spectrum was given by

$$\hat{S} e^{-t_{P-1}(1-\alpha/2)\hat{\sigma}} < S \leq \hat{S} e^{t_{P-1}(1-\alpha/2)\hat{\sigma}}. \tag{14}$$

### Coherence and phase estimation

To jackknife coherence and phase spectral estimates, we assumed the availability of $P$ complex transform pairs, $x_k(f)$ and $y_k(f)$. We defined the delete-one estimates of the coherency as

$$\hat{c}_{\setminus j} = \frac{\sum\limits_{\substack{k=1 \\ k \neq j}}^{P} x_k(f) y_k^*(f)}{\left[ \sum\limits_{\substack{k=1 \\ k \neq j}}^{P} |x_k(f)|^2 \sum\limits_{\substack{k=1 \\ k \neq j}}^{P} |y_k(f)|^2 \right]^{1/2}} \tag{15}$$

for $j = 1, 2, \ldots, P$, plus the standard estimate with nothing omitted (Thomson and Chave, 1991). From these, we transformed to the almost normal variates

$$Q_{\setminus j} = \sqrt{2P - 2} \tanh^{-1}\left( \left| \hat{c}_{\setminus j} \right| \right) \tag{16}$$

and obtained estimates and tolerances of the coherence spectrum.

Delete-one phase factors were constructed as

$$e_{\setminus j} = \frac{\hat{c}_{\setminus j}}{\left| \hat{c}_{\setminus j} \right|}, \tag{17}$$

with an average value

$$e_{\cap} = \frac{1}{P} \sum_{j=1}^{P} e_{\setminus j}. \tag{18}$$

This approach provided an estimate of the phase variance as

$$V\left\{ \hat{\phi} \right\} = 2(P - 1) \cdot \left( 1 - |\phi_{\cap}| \right), \tag{19}$$

in which $\phi_{\cap} = \arg\{ e_{\setminus j} \}$, an application of the jackknife method to standard phase statistics (Fisher, 1993). In particular, the estimate took periodicity into account and was equivalent to

$$\frac{P - 1}{P} \sum_{j=1}^{P} \left( \phi_{\setminus j} - \phi_{\cap} \right)^2, \tag{20}$$

in which $\phi_{\setminus j} = \arg\{ e_{\setminus j} \}$ for small phase dispersion.

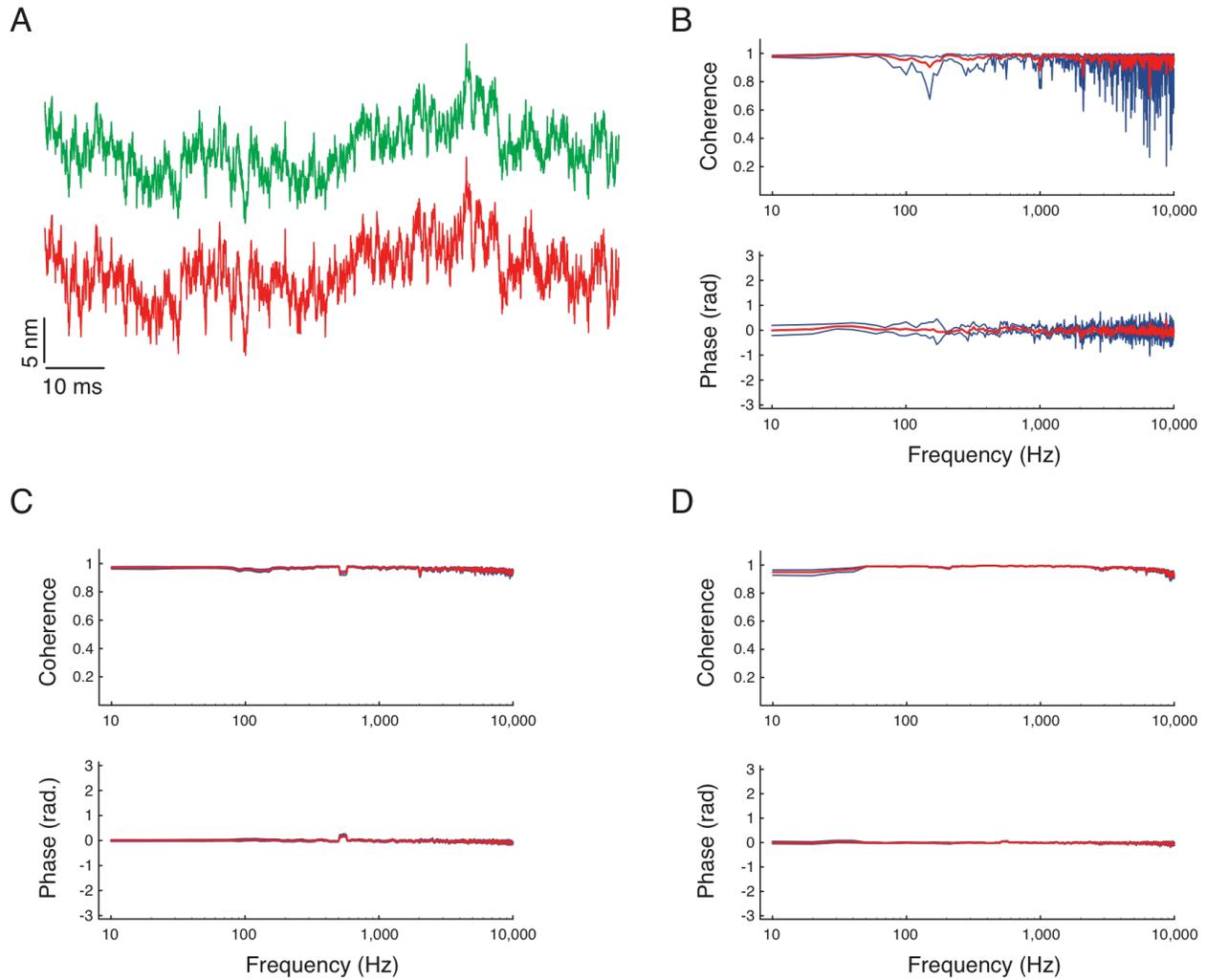

**Figure S1: Time series and coherency spectra from an auditory hair bundle.** (A) Simultaneous recordings with two interferometric beams show highly similar patterns of motion at the opposite edges of a gecko's hair bundle. (B) The coherence and phase spectra for the records in panel (A) are shown in red, together with their associated 95 % confidence intervals in blue. (C) Averaging the coherence and phase spectra from 20 records for the opposite edges of the same hair bundle greatly reduces the experimental uncertainty. The confidence intervals were obtained by jackknifing the entire set of records. (D) The average spectra from 20 records with the two beams positioned at an identical position in the same hair bundle closely resemble those in panel (C).

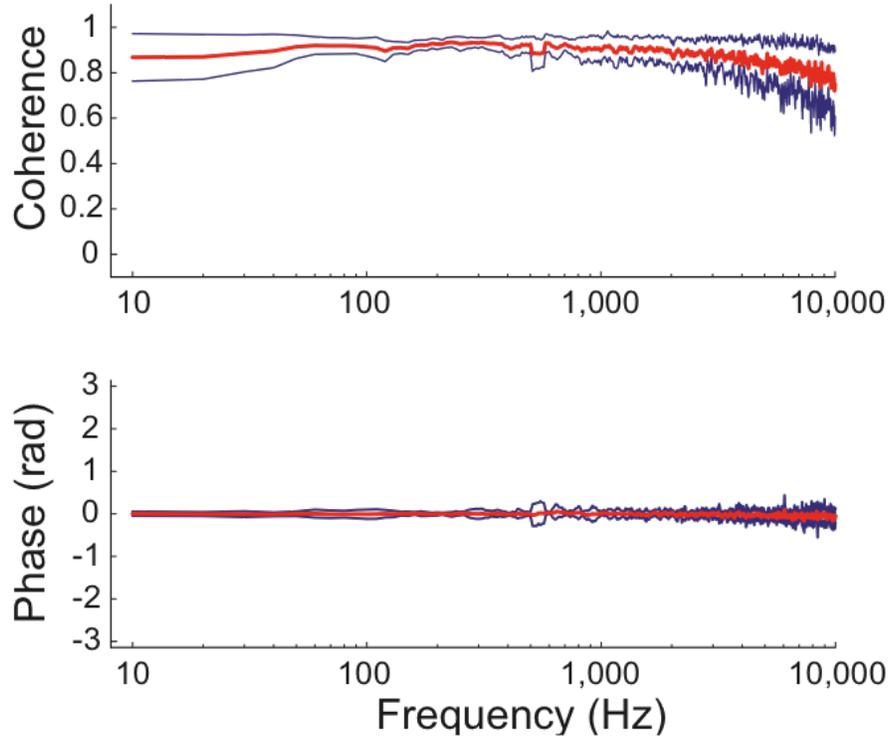

**Figure S2: Coherence and phase spectra from auditory hair bundles.** The figure displays the average values (red) and the associated standard deviations (blue) of the coherence and phase spectra for recordings from the opposite edges of five hair bundles from the gecko's basilar papilla.